# Ferrofluid bend channel flows for multi-parameter tunable heat transfer enhancement – Part 2: Deep Learning and Neural Network Modeling

Authors: Nadish Anand[a,1], Prashant Shukla[b], & Warren Jasper[a]

## Abstract

This work is the second in a series focused on ferrofluid bend channel flows. Here, ferrofluid flows in bend channels are modeled using machine learning methods, based on data generated from the CFD simulation discussed in the first work in this series. Predicting convective heat transfer in ferrofluid flows influenced by magnetic fields is key to advancing thermal management in microscale and energy-intensive systems. In this study, we introduce a data-driven framework to estimate four region-specific Nusselt numbers ($Nu_c, Nu_b, Nu_{b1}, Nu_{b2}$) from high-fidelity CFD simulations of ferrofluid flow through a bend channel, with different parameters that can tune the flow. These outputs are modeled as functions of seven relevant input features (flow tuning parameters): ferrofluid volume fraction ($\phi$), outer radius of the bend ($R_o$), distance between the bend center and current-carrying conductors ($dist_1$), angle between wires and horizontal ($\alpha$), current in the two current-carrying conductors ($I_1$ and $I_2$ ), and flow Reynolds number ($Re$). All these parameters were extensively varied in CFD simulations to characterize the flows. We trained and compared fully connected neural networks (NN), XGBoost, and Random Forest regressors. Our optimized NN achieves strong predictive performance on the complete dataset of 15,876 parametric variation cases. Beyond accuracy, the study emphasizes model transparency and reliability: using Permutation Importance and SHAP to interpret feature contributions, applying Monte Carlo Dropout to quantify prediction uncertainty, and conducting ablation and residual analyses to assess robustness. Results demonstrate that neural networks, when combined with physics-informed features and modern interpretability tools, provide a trustworthy and scalable surrogate for magnetically driven thermofluid systems. This framework is adaptable to broader applications in heat exchanger optimization, smart cooling devices, and the design of magnetic nanofluids.

## 1. Introduction

Effective prediction of convective heat transfer is essential for designing and optimizing advanced thermal systems, particularly in applications such as microelectronics cooling, energy harvesting, biomedical devices, and aerospace propulsion. The Nusselt number is a key parameter in assessing these systems. Recently, ferrofluids, which are colloidal suspensions of magnetic nanoparticles, have gained notable interest because of their adjustable thermal properties when subjected to magnetic fields[1–6]. Modeling heat transfer in ferrofluid flows remains challenging due to the complex interplay among magnetic, inertial, and geometric effects, leading to highly nonlinear and spatially intricate flow patterns [7–10]. Although traditional methods, such as computational fluid

a – North Carolina State University, b – Royal Melbourne Institute of Technology
1 – Corresponding Author: nanand@alumni.ncsu.edu

dynamics (CFD), provide accurate estimates, they are often too computationally intensive for real-time simulations, iterative design processes, or uncertainty analysis over extensive parametric ranges.

The application of machine learning (ML) techniques in thermal system modeling has gained traction over the past decade, particularly in scenarios where traditional simulations, such as computational fluid dynamics (CFD), are computationally intensive[11,12]. A growing body of work has demonstrated the potential of data-driven models to predict heat transfer behaviors in various fluid contexts, including nanofluids, magnetized flows, and geometrically complex domains. Khosravi and Malekan[13] trained artificial neural networks to estimate the convective heat transfer coefficient in $Fe_3O_4$ - water ferrofluids under varying magnetic field conditions, leveraging CFD-generated datasets. Their work demonstrated that ML could approximate CFD outputs at significantly lower computational cost. Similarly, Rehman et al.[11] employed ML models to study heat transfer in tangent hyperbolic fluids flowing over magnetized surfaces with internal heat generation, confirming the suitability of data-driven methods for non-Newtonian fluid systems. Abhijith and Soman [14] provided a comprehensive review of ML-based modeling techniques for nanofluid flows, with particular emphasis on compact heat exchangers and electronic cooling. Ullah et al. [15] discussed hybrid frameworks that combine CFD, experimental data, and machine learning to predict thermophysical properties of nanofluids, advocating for their use in industrial process optimization. Pai and Banthiya [16] proposed a transfer-learning-based surrogate model to estimate nanofluid thermal conductivity across multiple systems, highlighting the value of pre-trained models for generalized heat transfer predictions. Similarly, Durgam and Kadam [17] demonstrated the predictive power of combining analytical correlations with ML models for both viscosity and thermal conductivity using structured datasets. Similarly, Esmaeli et.al.[18] developed a study on ferrofluid flow-based heat transfer enhancement under an alternating magnetic field, experimentally measuring the convective heat transfer enhancement in heated ferrofluid straight-channel flows as different ferrofluid chemistry parameters were varied, and developed machine learning models to predict the same.  Finally, Ramezanizadeh et al. [19] Conducted a focused review on the use of intelligent systems, including support vector machines (SVM), artificial neural networks (ANN), and hybrid ensembles for predicting thermal conductivity in nanofluids. Their study reinforces the growing consensus that ML-based surrogates can serve as viable alternatives to physics-based solvers under appropriate constraints.

The rapid rise of machine learning (ML) has transformed how researchers approach complex thermal–fluid problems, especially those involving highly nonlinear interactions. Modern deep learning methods are now capable of discovering intricate patterns in data that traditional modeling techniques often struggle to capture. When these models are paired with explainable AI tools, they not only predict outcomes with remarkable accuracy but also illuminate the underlying physics that govern system behavior.

To overcome these limitations, this work introduces a comprehensive ML framework tailored for ferrofluid flow in a bent channel under the influence of external magnetic fields. Rather than restricting the analysis to a single thermal metric or simplified geometry, our study advances the state of the art by:

- Predicting four physically relevant Nusselt numbers corresponding to distinct spatial regions of the bent ferrofluid channel, enabling a multi-output understanding of heat transfer behavior under magnetic modulation.

- Leveraging a diverse and high-dimensional set of magnetic, geometric, and flow features, ensuring that the surrogate model generalizes across realistic ferrohydrodynamic operating conditions instead of narrow or low-variance datasets.

- Training and comparing neural networks, XGBoost, and Random Forests on a curated high-fidelity simulation dataset, allowing us to assess how different ML architectures capture nonlinear interactions between magnetic forcing, channel curvature, and thermal transport.

- Employing both global (Permutation Importance) and local (SHAP) interpretability frameworks to explain how individual physical parameters influence the predicted Nusselt numbers, which helps convert the model from a black box into a transparent analytical tool.

- Assessing predictive reliability through Monte Carlo Dropout-based uncertainty quantification, allowing the surrogate to express confidence bounds and distinguish between high and low certainty predictions.

- Validating robustness through feature ablation studies and residual error distribution analysis, which reveal how model performance changes when influential features are removed, perturbed, or sparsely represented.

By integrating these components, the framework delivers a scalable, interpretable, and physically grounded surrogate model that predicts magnetically tuned ferrofluid heat-transfer behavior with engineering-grade reliability. This unified structure, which combines multi-output prediction, high-dimensional feature learning, transparency, uncertainty estimation, and robustness evaluation, moves well beyond prior literature and provides a practical tool for applications where thermal performance and magnetic field control must be optimized together.

This comprehensive modeling effort aims not only to provide predictive accuracy but also to establish scientific credibility and ensure physical interpretability, thereby facilitating the adoption of machine learning in thermofluidic systems. In the following section, we briefly discuss the Bend channel flow CFD simulations and the available datasets for machine learning modeling. In the third section, we describe the neural network architecture and the evaluation criteria used to assess its performance. In the fourth section, we present the neural network results and evaluate its performance using primary, secondary, and tertiary evaluation methods.

## 2. CFD Simulations and Dataset

The data used to train neural networks come from CFD simulations of bend-channel flows, which are discussed in detail in the first article of this series. The simulations model a ferrofluid flow in a bend channel, with the bend subjected to an external magnetic field generated by two current-carrying conductors. The simulations span a parametric space of seven parameters, covering system configuration, geometry, ferrofluid chemistry, and flow conditions. The channel is also subjected to a uniform heat flux on the walls, and the resulting heat transfer enhancement from the introduction of an external non-uniform magnetic field is characterized by four region-specific Nusselt numbers.

## Overview of CFD Simulations

To summarize the geometry, **Figure 1** and **Figure 2** represent the system configuration for the elbow channel with an outer radius of 2 channel widths from the bend center point [1].

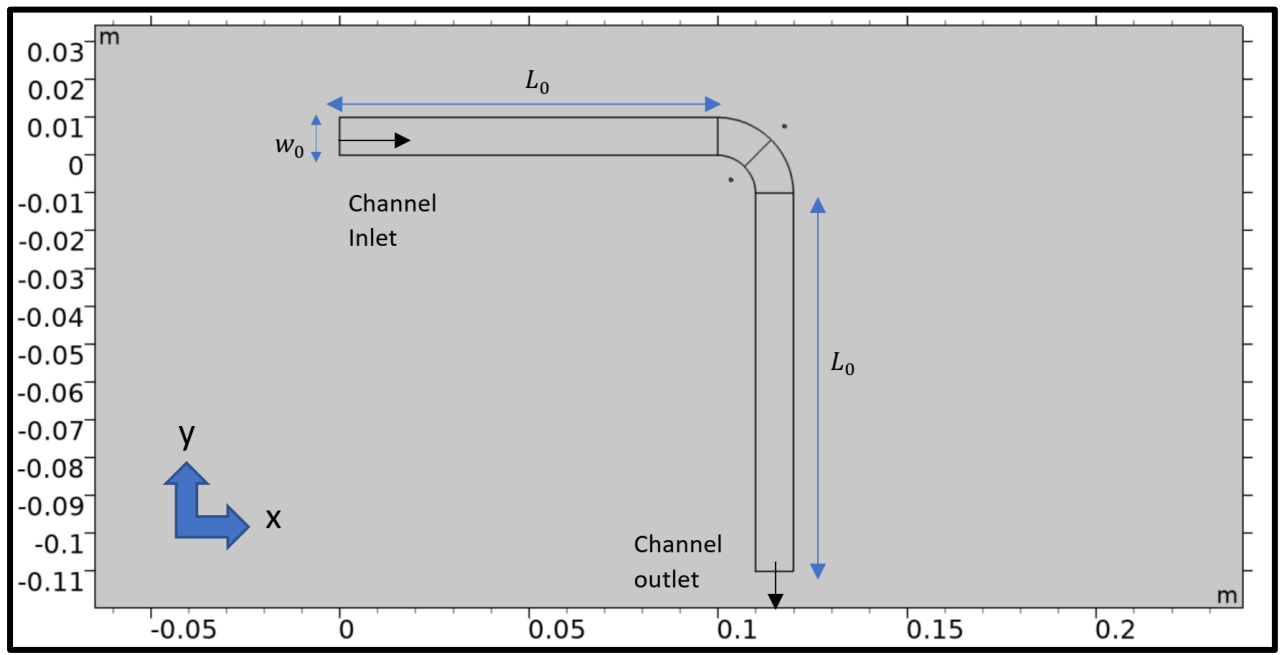

**Figure 1** Configuration of the elbow channel

The lengths of the two straight portions of the channel before and after the bend is $L_0$, which is set to 10 channel widths, leading to a hydrodynamically fully developed flow before it enters the bend. The wires can carry current in both directions, i.e., into and out of the plane. In **Figure 2** the parametric placement of the wires is shown along with the bend inlet, outlet, and the bend center. The bend center is chosen such that it divides the bend and, by extension, the channel into two parts of equal area, separated by an angle of 45°. Another variation is within the angle that the line joining the two wires makes with the horizontal. **Figure 3** shows one of the three different cases of symmetrical angular displacement (for $\alpha = 30°$) of the wires with respect to both the channel centers. The wires are always kept at a distance of $dist_1$ from the center of the bend and are given an angular displacement of $\alpha = 30°\ to\ 60°$ in 5° increments resulting in a distinct simulation case each time the angle changes. To illustrate the unique flow fields obtained from the tuned ferrofluid flow configurations, **Figure 4** shows the velocity magnitude of the ferrofluid in the bend, for a $Re = 15$, $I_1 = -5\,A$, $I_2 = -5\,A$, $\alpha = 45°$, $dist1 = w_0$, $rad_{out} = 2\,w_0$ and $\phi = 0.05$. The intricate flow patterns because of the external magnetic field gives rise to heat transfer enhancement, which changes with changing system parameters, and varies globally and locally.

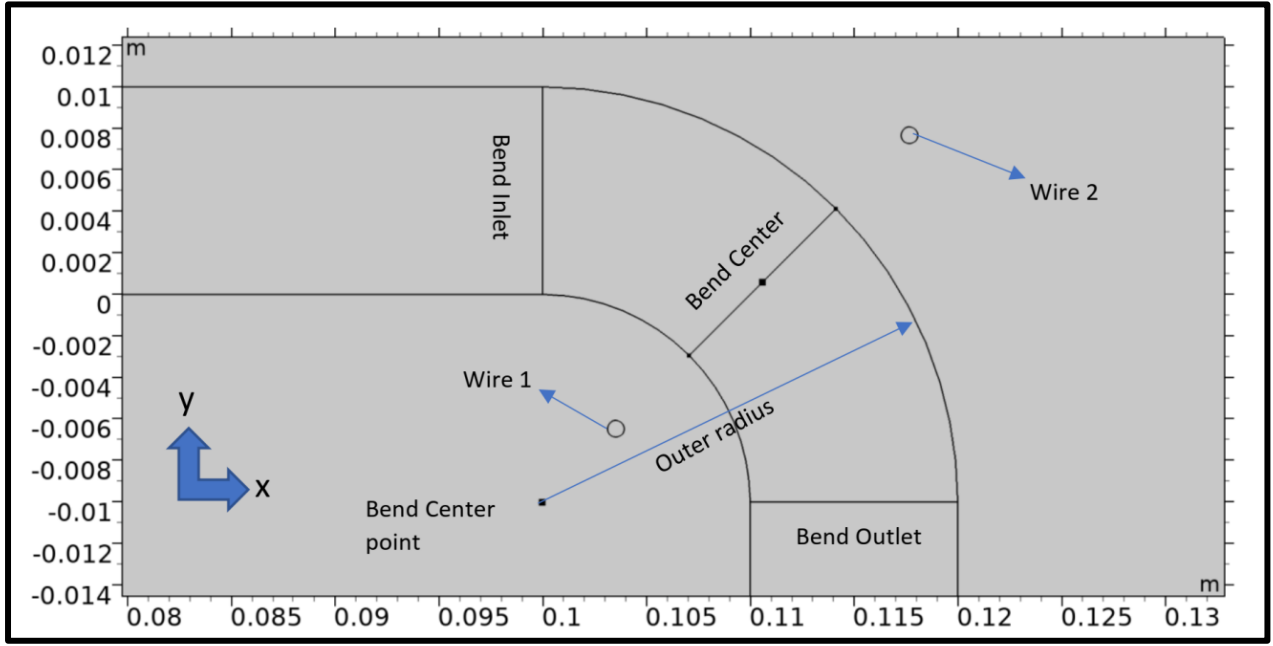

**Figure 2** Elbow channel bend for outer radius of 2 channel widths

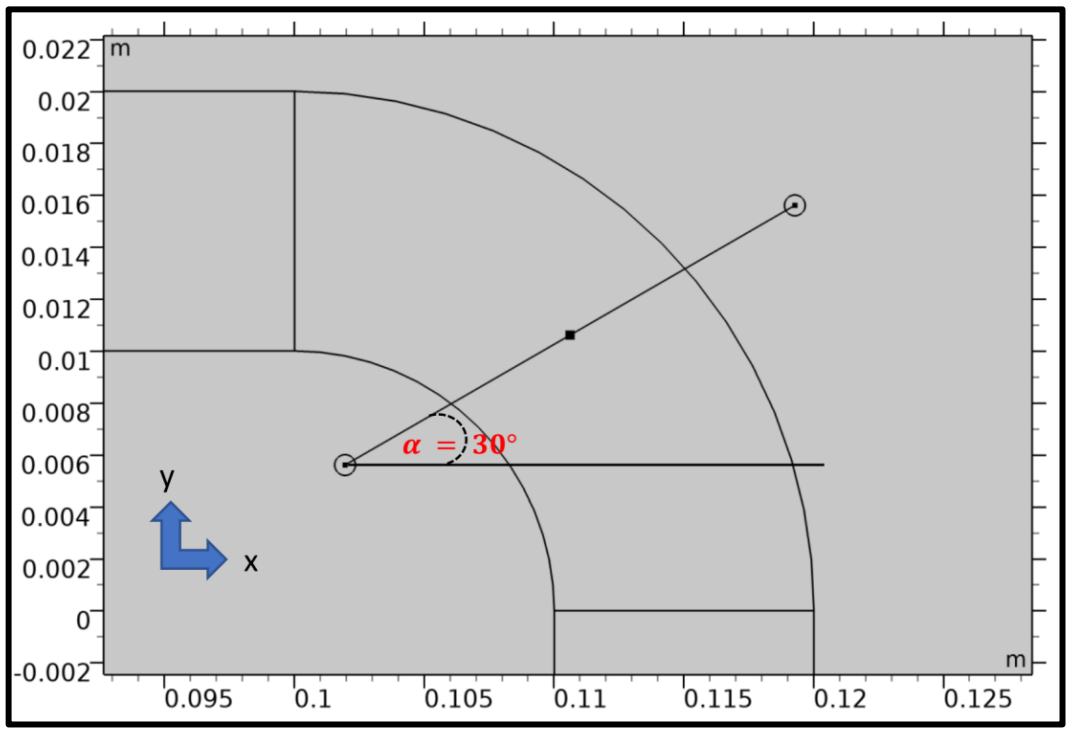

**Figure 3** Figure showing angle of wires with the horizontal $\alpha = 30°$ for channel 1

Table 1 Parameter varied for simulation and their values/limits

| Parameter | Values/Limits |
| --- | --- |
| Current in wire 1 ($I_1$) | -5, 0 and 5 [A] |
| Current in wire 2 ($I_2$) | -5, 0 and 5 [A] |
| Distance between wires and bend center ($dist_1$) | $0.75w_0$ & $w_0$ |
| Outer Radius of the bend ($R_o$) | $2w_0$ (channel 1), $4w_0$ (channel 2) and $6w_0$ (channel 3) |
| Angle of wires with horizontal ($\alpha$) | 30° to 60°, in 5° advancements (7 values) |
| Reynolds Number ($Re$) | 5 to 25 |
| Ferrofluid concentration ($\phi$) | 5% and 10% |

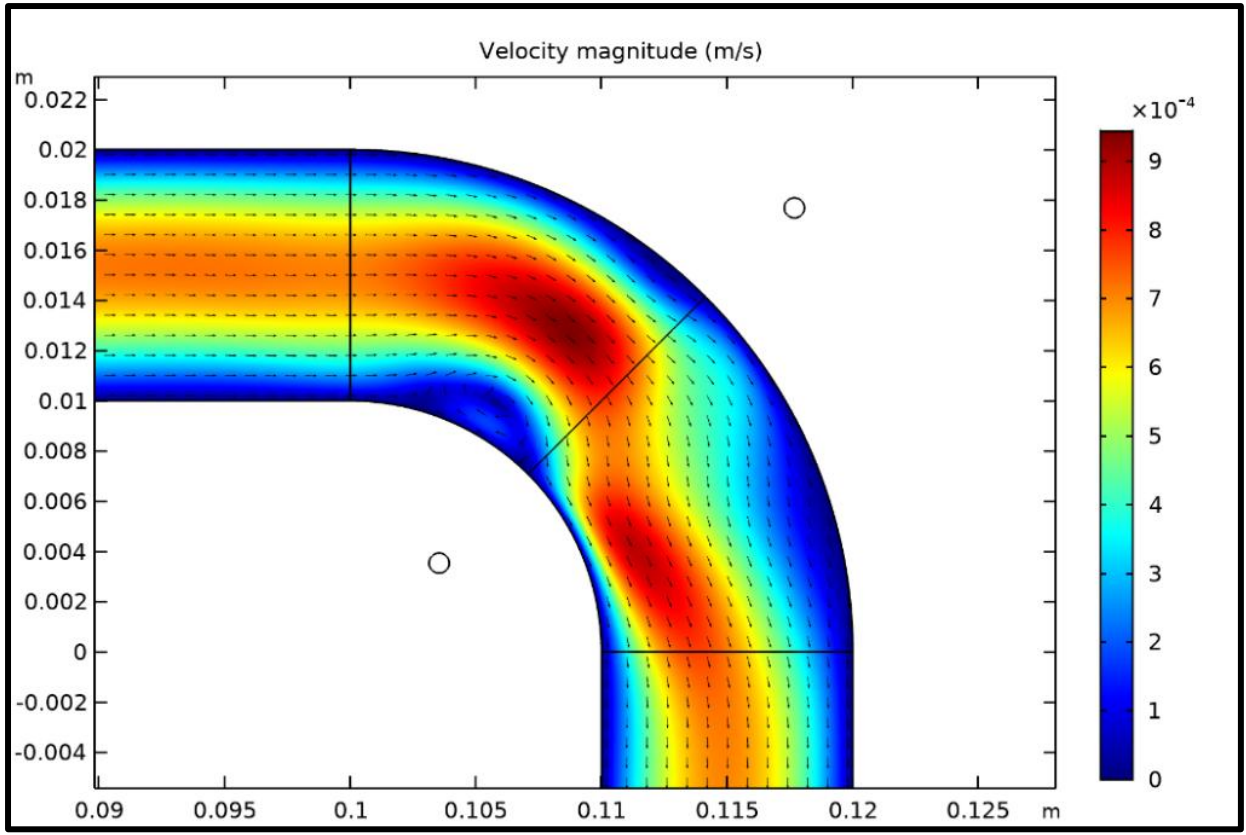

**Figure 4** Velocity Magnitude in the bend contour plot for one of the 15876 cases

## Dataset

The dataset used in this study was created from high-precision CFD simulations of the configurations discussed in the previous section. The input parameters for the Neural Networks are listed in **Table 1** Parameter varied for simulation and their values/limits and are the same as those used in the CFD simulations. The total number of parametric variations is 15876.

The goal of this study is to predict the following four distinct Nusselt number outputs, each corresponding to a physically meaningful region in the ferrofluid domain:

- $\boldsymbol{Nu_c}$: Average Nusselt number for the whole channel
- $\boldsymbol{Nu_b}$: Average Nusselt number for the bend
- $\boldsymbol{Nu_{b1}}$: Nusselt number for the first bend section
- $\boldsymbol{Nu_{b2}}$: Nusselt number for the second bend section

All input features were z-score normalized, a common preprocessing technique that improves neural network convergence by centering and scaling feature distributions. The final dataset, curated to focus on convective regimes of interest, consisted of 15,876 samples. This dataset captures the coupled effects of geometry, magnetic field strength, and flow inertia, offering a rich landscape for machine learning models to learn complex nonlinear mappings between input parameters and thermal response.

## 3. Neural Network Architecture

Our final neural network architecture was designed to strike a balance between capacity and generalization, while capturing the complex nonlinearities inherent in magnetically driven thermal flows. The neural network has the following layers and features:

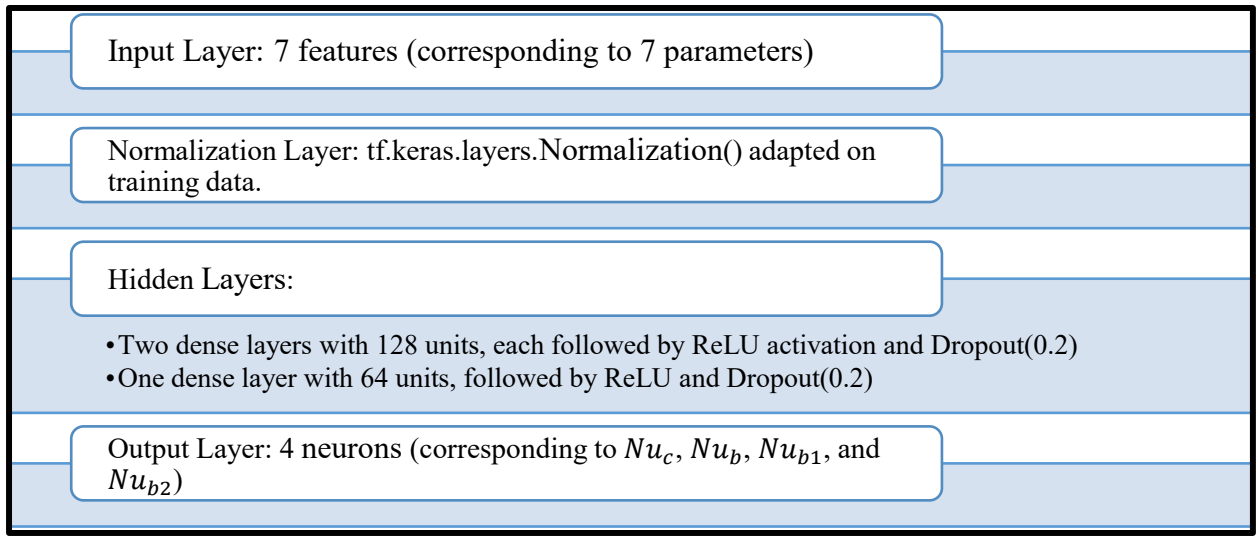

**Figure 5** Neural Network Configuration & Features

The model was trained using the Adam optimizer with an initial learning rate of 0.001. The loss function was Mean Squared Error (MSE), with Mean Absolute Error (MAE) tracked as a secondary metric. Training was conducted for up to 150 epochs, using the parameters shown in **Figure 6**. These parameters allow for effective execution of the NN.

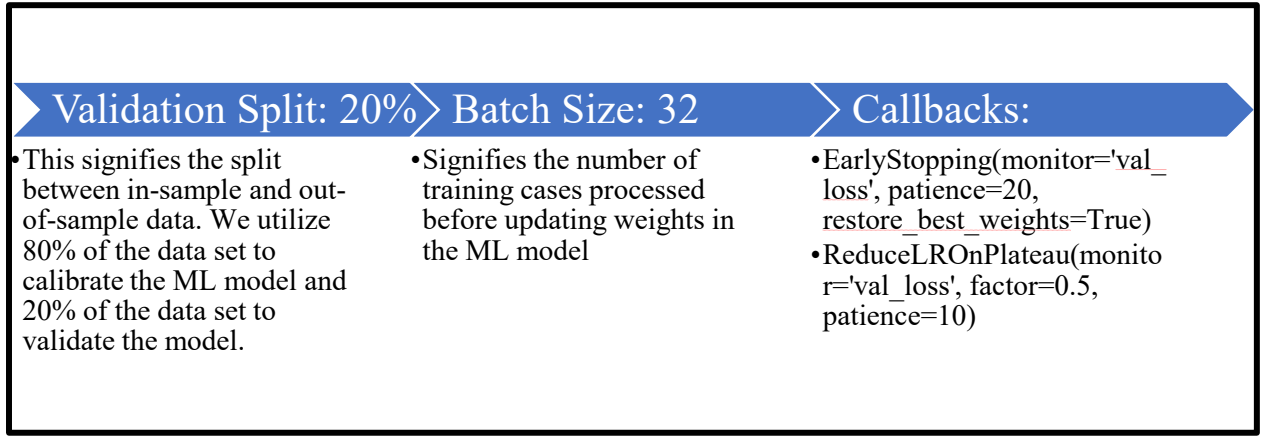

**Figure 6** Neural Network parameters

This architecture yielded the best performance among the evaluated models, particularly in terms of generalization and robustness across all four Nusselt number target outputs.

## 3.1 Methodology

To evaluate the predictive performance of different machine learning models for heat transfer estimation in ferrofluid flows, we employed three distinct regression approaches:
- Fully Connected Neural Network (NN): A flexible, data-driven method capable of capturing complex non-linearities in high-dimensional input spaces.
- XGBoost Regressor**:** A high-performance, tree-based ensemble model known for its scalability, regularization, and ability to handle structured data effectively [20].
- Random Forest Regressor**:** An ensemble of decision trees offering robust performance and reduced variance through averaging and bagging [21].

The dataset was partitioned into 80:20 train-test splits, with all models trained and evaluated on identical subsets to ensure fair performance comparisons. To optimize the neural network, we monitored the validation loss and employed adaptive learning rate scheduling with "ReduceLROnPlateau", along with "EarlyStopping" for regularization and overfitting mitigation.

## 3.2 Evaluation Metrics

Performance was measured using three standard regression metrics:

- Mean Absolute Error (MAE):

$$MAE = \frac{1}{n}\sum_{i=1}^{n}|y_i - \hat{y}_i|$$

- Root Mean Square Error (RMSE):

$$RMSE = \sqrt{\frac{1}{n}\sum_{i=1}^{n}(y_i - \hat{y}_i)^2}$$

- Coefficient of Determination (R²):

$$R^2 = 1 - \frac{\sum_{i=1}^{n}(y_i - \hat{y}_i)^2}{\sum_{i=1}^{n}(y_i - \bar{y})^2}$$

Here, $y_i$ are true values, $\hat{y}_i$ are predicted, and $\bar{y}$ is the mean of actual values. $MAE$ quantifies average error magnitude, $RMSE$ highlights sensitivity to outliers, and $R^2$ gauges predictive alignment with ground truth.

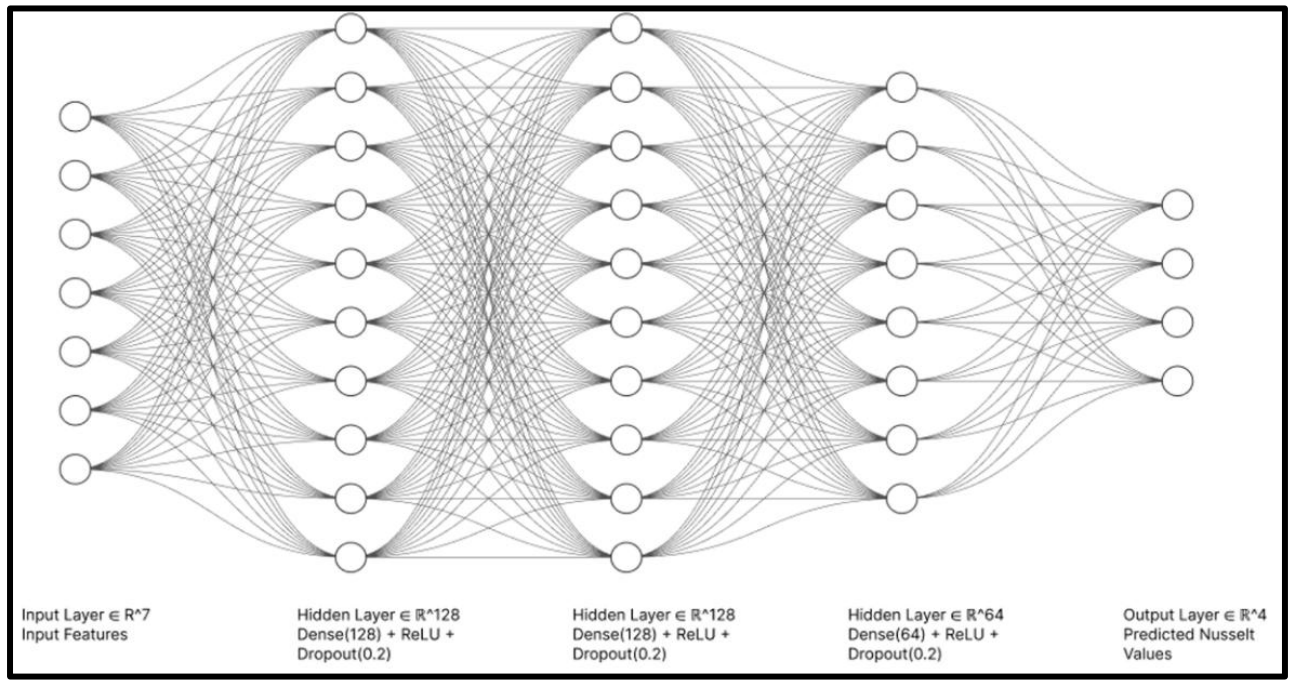

**Figure 7** Neural Network architecture visualized

## 4. Results and Discussions

In this section we will first discuss the results from the Neural network and measure its performance based on the metrics we discussed in the previous section and compare them with those from established methods like XGBoost and Radom Forests. From there we will evaluate performance of the neural network by using Advanced performance evaluation through secondary and tertiary methods like Permutation Importance, SHAP analysis, Uncertainty Quantification, etc.

### 4.1 Traditional Predictive Performance Evaluation methods

We assessed the predictive accuracy of the trained models using standard regression metrics, specifically Root Mean Squared Error (RMSE), Mean Absolute Error (MAE), and Coefficient of

Determination ($R^2$), as described in Section 3.2. These metrics were computed on a held-out test set consisting of 20% of the dataset (3,175 samples).

Neural Network Performance

The optimized neural network demonstrated consistently strong performance across all four target variables, as shown from the metric values in **Table 2**.

Table 2 Neural Network performance metrics on the test set

| Metric | $Nu_c$ | $Nu_b$ | $Nu_{b1}$ | $Nu_{b2}$ |
|---|---|---|---|---|
| RMSE | 0.0452 | 0.1178 | 0.1490 | 0.1553 |
| MAE | 0.0356 | 0.0583 | 0.0753 | 0.0887 |
| $R^2$ | 0.8319 | 0.9779 | 0.9724 | 0.9743 |

These results confirm that the model accurately captures the underlying relationships for $Nu_b$, $Nu_{b1}$, and $Nu_{b2}$, achieving R² values exceeding 0.97. The slightly lower R² for $Nu_c$ likely stems from reduced feature variance and weaker sensitivity to dominant input parameters such as Reynolds number ($Re$) and the two currents ($I_1, I_2$). This directly conforms with the underlying physics of the setup, where a relatively small area of the whole channel (the bend) is subjected to flow manipulation by changing different geometric, chemical, and electromagnetic parameters.

Prediction vs Actual Plots

To further assess the predictive fidelity of our neural network, we generated scatter plots shown in **Figure 8** comparing the predicted and actual values for each of the four Nusselt number outputs: $Nu_c, Nu_b, Nu_{b1}$, and $Nu_{b2}$. Such plots are a standard tool in regression model validation, offering an intuitive means to visually evaluate model calibration, systematic bias, and variance patterns across the output domain [22,23]. The plots reveal a tight clustering of data points along the ideal diagonal line ($y = x$) for $Nu_b, Nu_{b1}$, and $Nu_{b2}$, confirming high model calibration and minimal residual spread in these regions. This strongly supports the quantitative results presented earlier, where these outputs achieved $R^2$ values above 0.97 and minimal RMSE. $Nu_c$ on the other hand, exhibits a slightly wider dispersion around the diagonal, indicating that this output remains more challenging to predict. This observation aligns with insights from both Permutation Importance (PI) and SHAP analysis $Nu_c$ is relatively less influenced by dominant features such as Re and the currents ($I_1, I_2$) and may depend more on subtle or higher-order interactions not fully captured by the selected inputs. Additionally, its lower variance in the dataset inherently reduces the regression signal-to-noise ratio, making perfect predictions less attainable [24].

Importantly, the scatter plots do not display evidence of heteroscedasticity (i.e., increasing prediction error with output magnitude), nor do they indicate systematic under- or over-prediction across the range apart from minor underestimations in high-gradient flow conditions. These regions correspond to Higher Nusselt number regimes, which are invariably in flow configurations where the Reynolds Number is low, showcasing the inverse relationship between flow inertia and magnetic field effects on the ferrofluid [25]. Together, these plots visually confirm the model's strong generalization ability.

### Baseline Comparisons: XGBoost and Random Forest

To contextualize the neural network's performance, we trained and evaluated XGBoost and Random Forest regressors using identical input features, splits, and preprocessing.

Table 3 Baseline Comparison with Random Forest & XGBoost

| Model | $Nu_c$ | $Nu_b$ | $Nu_{b1}$ | $Nu_{b2}$ |
|---|---|---|---|---|
| XGBoost R² Scores | 0.9742 | 0.9762 | 0.9712 | 0.9623 |
| Random Forest R² Scores | 0.9772 | 0.9489 | 0.9258 | 0.9715 |
| Neural Network R² Scores | 0.8319 | 0.9779 | 0.9724 | 0.9743 |

As evidenced in **Table 3** XGBoost demonstrated excellent predictive capacity, particularly for $Nu_c$, marginally outperforming the neural network in that regard. However, it lacked built-in support for multi-output regression and uncertainty quantification, limiting its scalability in broader simulation-to-surrogate pipelines. While Random Forest performed well for $Nu_c$ and $Nu_{b2}$, its predictive consistency for $Nu_b$ and $Nu_{b1}$ was comparatively weaker than XG Boost. This is likely due to its non-parametric nature and sensitivity to input dimensionality when capturing interactions in complex thermofluid systems.

Across all models, the neural network provided the most balanced and robust performance across local outputs, particularly excelling in scenarios with high nonlinearity and local flow transitions. While its performance suffered in predicting the global Nusselt number. Its extensibility to uncertainty-aware prediction and interpretability via SHAP and permutation importance, as demonstrated in the coming sections make it a reliable surrogate model for simulating Ferrofluid bend channel flows for heat transfer enhancement.

### 4.2 Advanced performance evaluation methods

Traditional analysis methods provide an overall assessment of model performance and establish the central tendency of model biases. However, traditional analysis methods do not provide deeper insights into the neural networks, such as the dependence of output predictions on different input parameters. This section introduces 4 new methods that go deeper into the internal workings of the designed neural network.

### 4.2.1 Permutation Importance (PI)

To investigate the contribution of each input feature to the predictive accuracy of our model, we employed Permutation Importance (PI), a robust, model-agnostic interpretability technique introduced by Breiman[21] and formalized further by Fisher et al.[26] It measures the change in model performance when a single feature's values are randomly shuffled, thereby breaking its relationship with the target variable. If the performance degrades significantly, the feature is deemed important.

In our case, where we aim to predict four Nusselt numbers simultaneously, PI enables us to quantify the sensitivity of each output to each input, grounded purely in model behavior rather than model weights or architecture.

Let f denote a trained model, and $D = \{(x_i, y_i)\}_{i=1}^{n}$ be the test dataset. Define a performance metric $M$ (here, $R^2$). The permutation importance of feature $j$ is:

$$PI_j = M(D) - M(D^{\pi(j)})$$

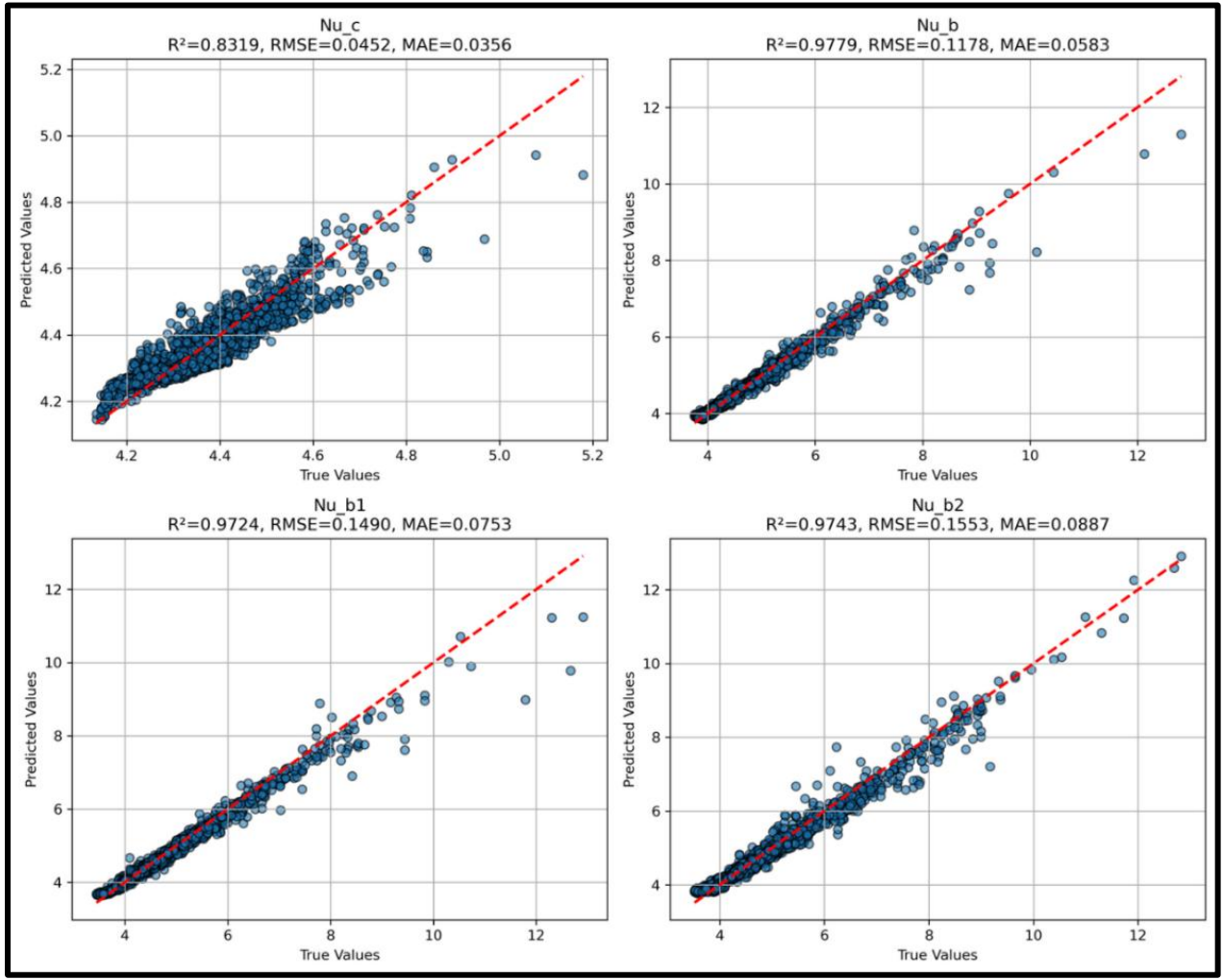

**Figure 8** Scatter plot for Prediction vs Actual Nusselt Numbers

where $D^{\pi(j)}$ is the dataset with the $j$-th feature randomly permuted across samples. A larger drop in $M$ implies greater importance. We repeated this process 5 times per feature to mitigate the randomness of single permutations and averaged the results. While modeling ferrofluid heat transfer through intricate geometries and subjected to non-uniform magnetic fields, it is crucial to understand which physical variables most significantly influence the model's predictions. This understanding helps validate that the machine learning model conforms to established physical principles, provides insights into the mechanisms involving magnetic and inertial contributions to heat transport, and guides future data collection and sensor design by highlighting key features.

Since we use a non-linear neural network, the direct interpretation of weights is not meaningful, making PI an ideal post-hoc diagnostic tool. We used the 'permutation importance()' function from Scikit-learn's 'inspection' module[27], applied separately for each of the four Nusselt numbers. The R² score was used as the performance metric. All experiments were conducted on the held-out test set, with 5 random permutations per feature. For detailed analysis of Permutation Importance **Table 6** (In Appendix A) provides the detailed values of feature-wise mean importance for all the Nusselt numbers. The permutation importance results are summarized below and visualized in **Figure 9** to **Figure 12**.

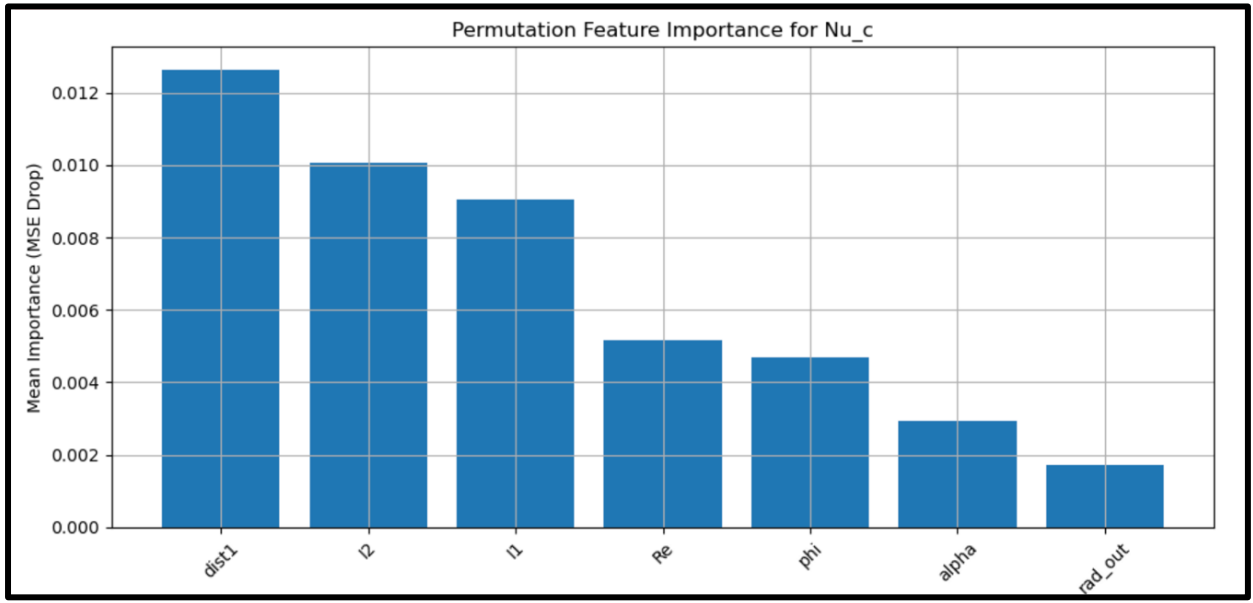

**Figure 9** Permutation importance for Nusselt number for the channel ($Nu_c$)

From **Figure 9** it is observed that $Nu_c$ is most sensitive to $dist_1$, $Re$, and the currents ($I_1, I_2$). This is expected as on a global level flow is governed by bulk fluid inertia and the Magnetic field, which results in enhancement of heat transfer over the baseline Nusselt number, when there is no Magnetic field present. From **Figure 10** $Nu_b$ is strongly influenced by $Re, dist_1$, and $R_o$.

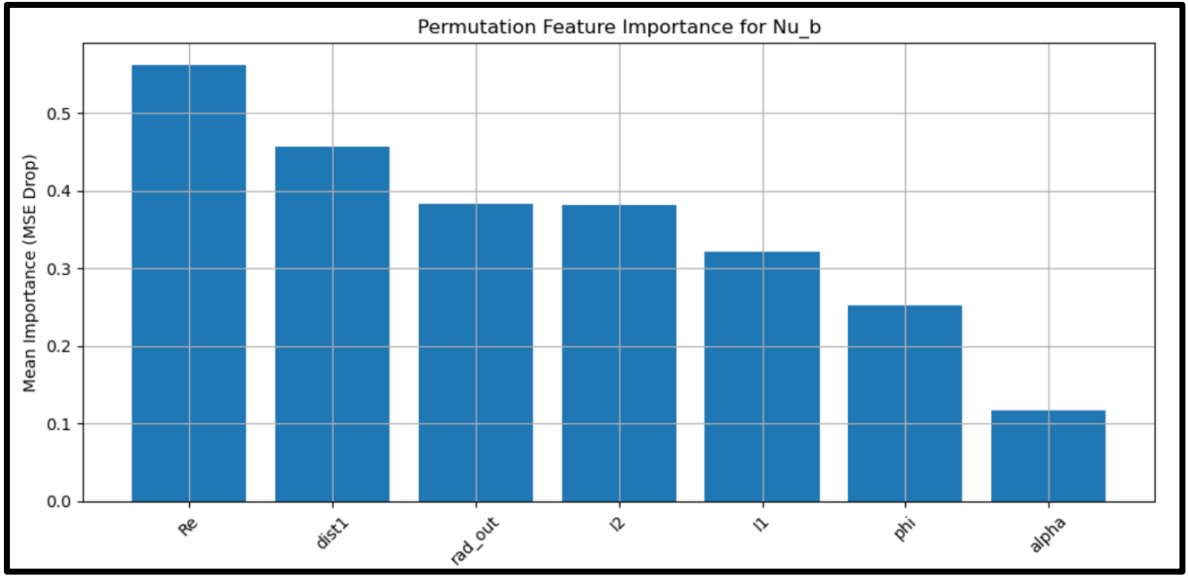

**Figure 10** Permutation importance for Nusselt number for the bend ($Nu_b$)

These features control the flow inertia; the strength of the magnetic field applied through the wires and the curvature of the bend. The currents ($I_1, I_2$) come after these three parameters, signaling the importance of all the local effects imposed on the flow, which tracks with the Physics. $Nu_{b1}$ & $Nu_{b2}$ showed distinct patterns: $Nu_{b1}$ emphasized $Re$, $I_2$, and $R_o$, while $Nu_{b2}$ leaned toward, $R_o$, and $Re$.

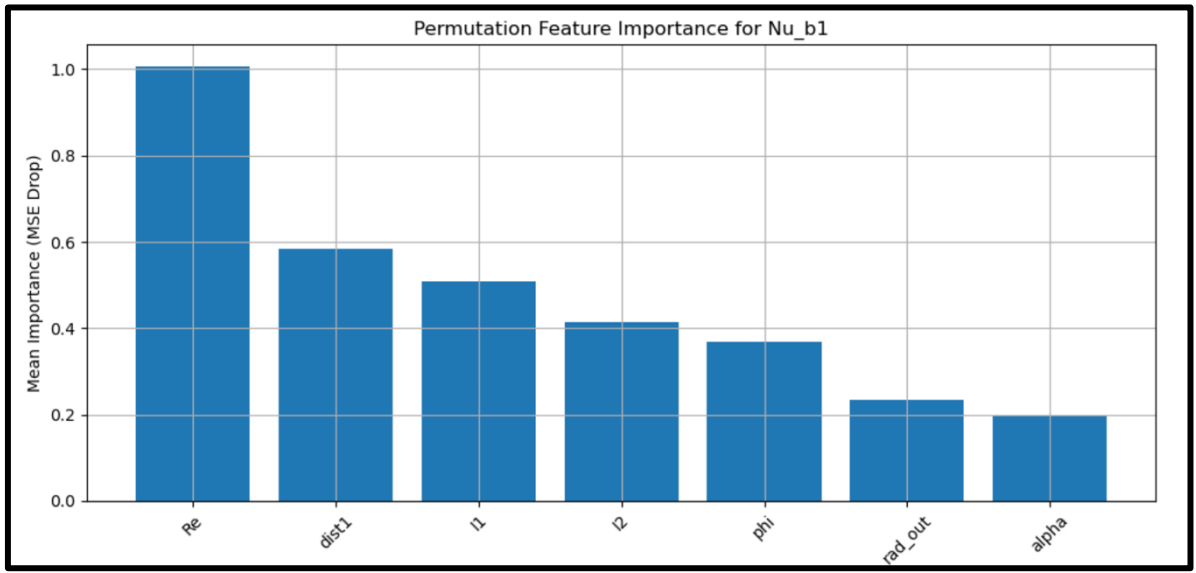

**Figure 11** Permutation importance for Nusselt number for the first bend section ($Nu_{b1}$)

These differences may reflect localized magnetohydrodynamic behaviors as the flow transitions around the curve.

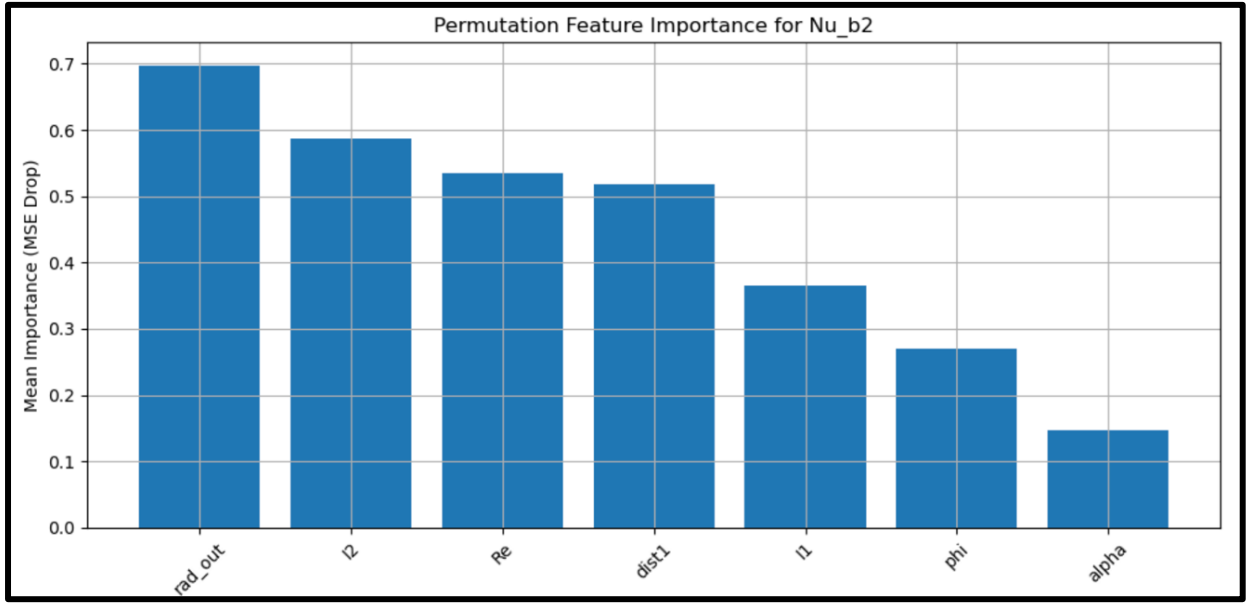

**Figure 12** Permutation importance for Nusselt number for the second bend section ($Nu_{b2}$)

The most influential feature across all outputs was $dist_1$, followed by $Re$ and the magnetic parameters. $\phi$ (volume fraction) and $\alpha$ (Angle of wires with horizontal) had marginal effects, consistent with their secondary role in altering local heat transfer under the studied conditions.

### 4.2.2 SHAP Analysis

To complement permutation importance and further enhance model interpretability, we employed SHAP (Shapley Additive Explanations)**,** a unified framework based on cooperative game theory introduced by Lundberg and Lee [28] that provides local explanations for individual predictions while remaining consistent and theoretically grounded.

SHAP assigns an importance value (positive or negative) to each feature for a given prediction, indicating how much the feature contributes to pushing the prediction higher or lower compared to the average output. Unlike permutation importance, which captures global impact by disrupting entire features, SHAP excels in revealing directionality, interactions, and distributional nuances at a per-sample level[25].

SHAP values are derived from Shapley values in cooperative game theory, which determine the fair payout to players (features) depending on their contribution to the coalition (model prediction). For a model $f(x)$, the SHAP value for feature $i$ is defined as:

$$\phi_i = \sum_{S \subseteq \mathcal{F} \setminus \{i\}} \frac{|S|!\,(|\mathcal{F}| - |S| - 1)!}{|\mathcal{F}|!} [f(S \cup \{i\}) - f(S)]$$

Where:

- $\mathcal{F}$ is the full set of features.
- $S$ is a subset of features not containing iii.

- $f(S)$ is the model's output when only features in $S$ are present.

In practice, we use "KernelSHAP" or "DeepSHAP" approximations for neural networks due to computational constraints[29]. While permutation importance reveals the average importance of a feature, SHAP provides instance-level insights, helping us understand how features influence predictions, whether positively or negatively. Recognize any nonlinear or context-specific interactions, such as when certain factors are only important at high levels. Ensure the model is reliable by comparing the overall PI with local SHAP values to verify consistency. We applied SHAP to our trained neural network model and generated summary beeswarm plots for each Nusselt number. **Figure 13** shows the summary plot from the SHAP analysis.

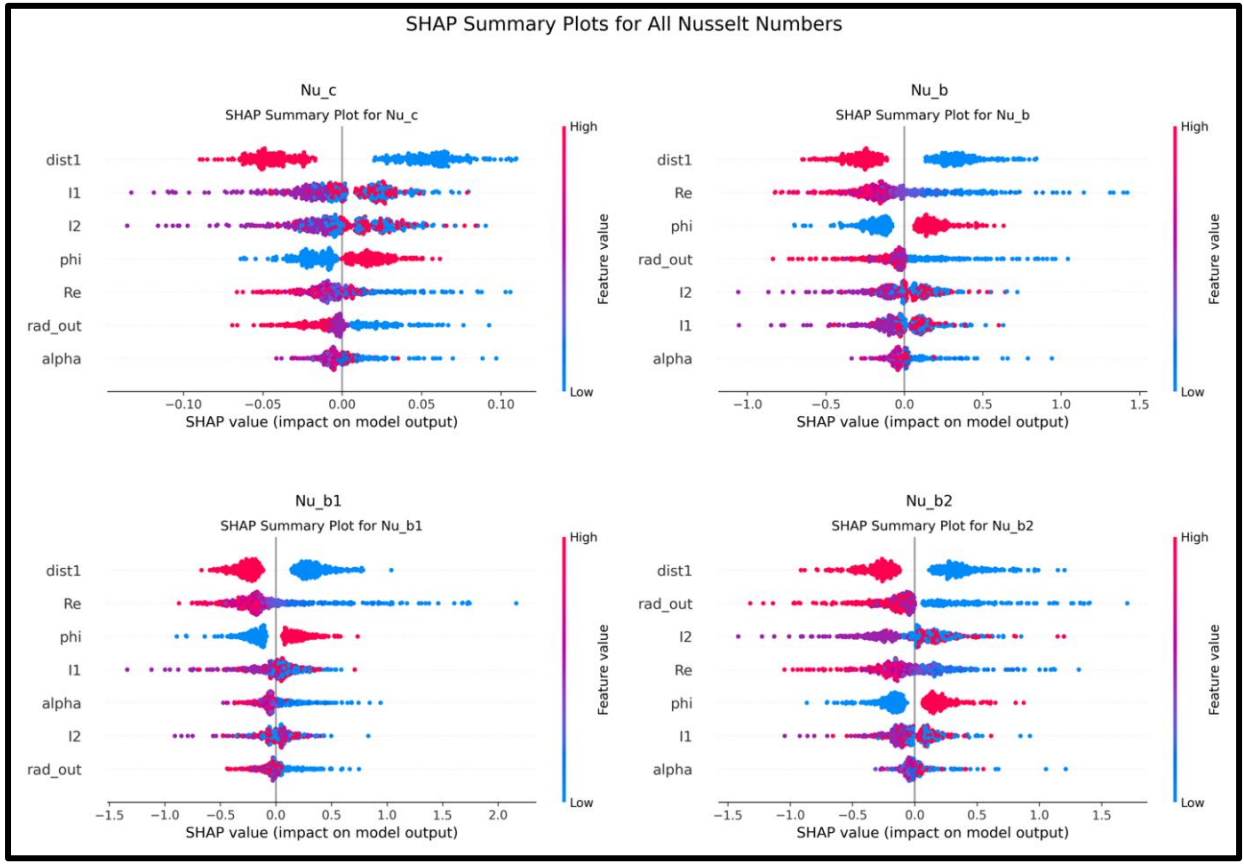

**Figure 13** SHAP Analysis Summary plot

The SHAP analysis basically is performed on the ML model acting on the 20% of the test data set, and then interprets how the model predictions of the 4 Nusselt numbers scale with the input parameters. If the scaling is consistent with physics, the model interpretability is valid. The SHAP beeswarm plots reveal a clear and physically consistent hierarchy of parameter influence across all four predicted Nusselt numbers. The magnetic field in the present study is generated by two current-carrying conductors oriented perpendicular to the flow plane, resulting in a strongly non-uniform magnetic field. The magnetic field strength scales with current and distance according to

$H = \frac{I}{(2\pi r)}$, while the magnetic body force acting on the ferrofluid scales with the gradient of the magnetic field squared $f_m \sim \mu_0 \chi \nabla(H^2)$, and the magnetic field gradient scales as $\nabla(H^2) \approx \frac{I^2}{r^3}$. Hence, magnetic tuning of the flow is governed primarily by spatial proximity and geometric configuration in addition to the current magnitude. This fundamental scaling explains the dominant role of the distance between the wires and the bend center ($dist_1$) across all Nusselt number predictions. This is correctly interpreted by the model, in this SHAP analysis

For the whole-channel averaged Nusselt number $Nu_c$, all seven input parameters influence heat transfer. The distance between the wires and the bend center ($dist_1$) dominates by controlling the magnitude of magnetic field gradients experienced by the flow. The Reynolds number $Re$ sets the inertial level of the flow, while the ferrofluid volume fraction $\phi$ determines magnetic susceptibility and thus the strength of magnetic body forces. The outer bend radius $R_o$ affects curvature-induced mixing, the wire angle $\alpha$ governs the alignment of magnetic forces with the flow direction, and the wire currents $I_1$ and $I_2$ scale the magnetic field intensity. However, due to spatial averaging over the entire channel, the influence of current magnitude is attenuated compared to geometric effects.

For the bend-averaged Nusselt number $Nu_b$, the interaction between flow inertia and magnetic forcing becomes more pronounced. The parameter $dist_1$ remains dominant due to its direct control of magnetic gradients within the bend. Reynolds number $Re$ strongly influences the development of curvature-induced secondary flows, which are either enhanced or suppressed by magnetic body forces. The ferrofluid concentration $\phi$ amplifies magnetic coupling through increased susceptibility, $R_o$ controls the smoothness of the flow curvature, $\alpha$ determines the directional projection of magnetic forces, and the currents $I_1$ and $I_2$ scale the overall Kelvin Body foce magnitude.

In the first bend section, represented by $Nu_{b1}$, magnetic effects are most effective at lower Reynolds numbers. The distance dist1 again governs the strength of magnetic gradients, while $Re$ determines whether magnetic forces can compete with inertial effects during boundary layer development. The ferrofluid concentration $\phi$ controls susceptibility-driven force amplification, $\alpha$ influences the redirection of developing secondary flows, $R_o$ affects curvature-induced flow development length, and the currents $I_1$ and $I_2$ scale the magnitude of magnetic forcing.

For the second bend section $Nu_{b2}$, the SHAP analysis indicates a magnetically sustained regime. The parameter $dist_1$ remains the primary control on magnetic gradients, while $\phi$ becomes more influential due to its direct scaling with magnetic susceptibility. The outer radius $R_o$ governs vortex persistence and wall interaction, $Re$ plays a reduced but non-negligible role, $\alpha$ determines the symmetry of magnetic forcing, and the currents $I_1$ and $I_2$ continue to scale the magnetic field strength sustaining heat transfer enhancement.

Overall, the SHAP beeswarm plots confirm that all seven parameters $dist_1$, $Re$, $\phi$, $R_o$, $\alpha$, $I_1$ and $I_2$ are physically meaningful and interdependent. The results demonstrate that magnetic tuning of ferrofluid heat transfer in bent channels is governed by gradient-driven magnetic forces rather than current magnitude alone, with distinct force balances operating in different bend regions. The machine learning model therefore captures region-specific ferrohydrodynamic physics.

### 4.2.3 Uncertainty Quantification

Uncertainty Quantification (UQ) has emerged as a critical component of modern machine learning workflows, particularly in physical modeling tasks where interpretability and risk-aware decision-making are essential. In this study, we employed Monte Carlo (MC) Dropout, a method formally introduced by Gal and Ghahramani (2016) [30], which casts standard dropout training as an approximate Bayesian inference technique. This allows for efficient estimation of model uncertainty without the computational burden of full Bayesian neural networks. The method has since become a cornerstone in UQ studies across physics, medical imaging, and engineering design, making it well-suited to our thermofluid modeling context.

During inference, MC Dropout involves retaining dropout layers in active mode and performing multiple stochastic forward passes through the network. This approach allows estimation of model uncertainty without the high computational cost associated with full Bayesian neural networks.

For each test sample, fifty stochastic forward passes were performed, yielding distributions of predicted Nusselt numbers for all four outputs. From these distributions, both the predictive mean and standard deviation were computed. The average predictive uncertainty across the test dataset is summarized in **Table 4** and visualized in **Figure 14**.

The whole-channel averaged Nusselt number, $Nu_c$, exhibited the lowest mean uncertainty. This behavior is physically expected, as $Nu_c$ represents a spatially averaged quantity that integrates heat transfer over both straight and curved sections of the channel. Local variations in magnetic forcing and flow structure are therefore partially averaged out, resulting in a more stable and predictable relationship between inputs and output.

In contrast, the bend-averaged Nusselt number, $Nu_b$, showed higher uncertainty than $Nu_c$. This reflects the increased complexity of heat transfer in the curved region, where curvature-induced secondary flows interact with spatially non-uniform magnetic forces. These coupled effects introduce stronger nonlinearities, making the prediction task more challenging.

The highest uncertainty was observed for the first bend-section Nusselt number, $Nu_{b1}$. This region corresponds to the entry portion of the bend, where boundary layers are still developing and curvature-induced vortices are being initiated. In this transitional regime, small changes in flow inertia, wire placement, or magnetic field strength can lead to significant changes in local heat transfer. As a result, the model exhibits greater uncertainty when predicting $Nu_{b1}$, reflecting genuine physical variability rather than poor model performance.

The uncertainty associated with the second bend-section Nusselt number, $Nu_{b2}$, was lower than that of $Nu_{b1}$ but higher than that of $Nu_c$. This trend is consistent with the downstream bend region, where secondary flows are more fully developed and magnetic forces act primarily to sustain or modify existing flow structures rather than initiate them.

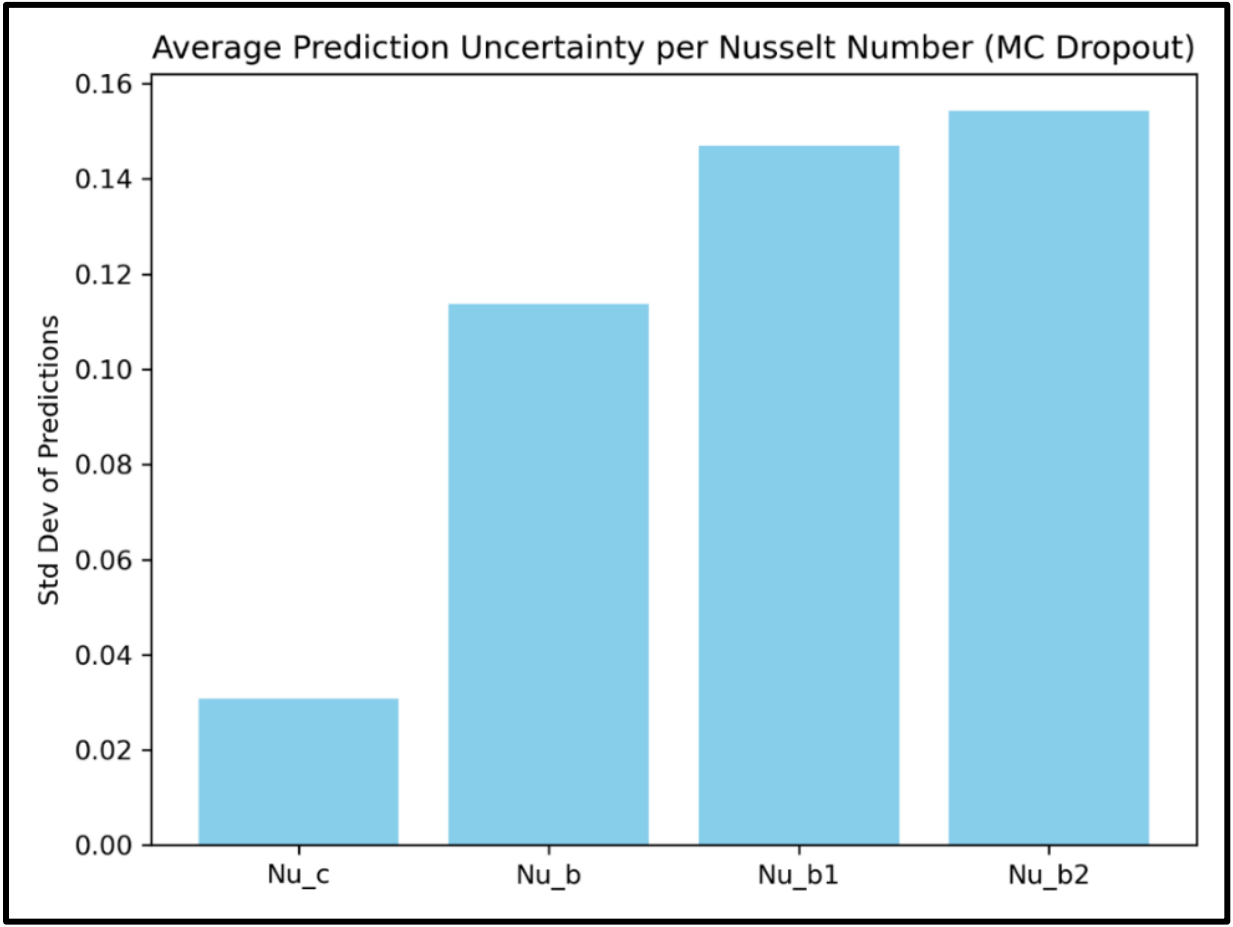

**Figure 14** Uncertainty quantification for prediction of Nusselt Number

**Table 4** Mean predictive uncertainty across target variables

| Output | Mean Std. Dev. |
|---|---|
| $Nu\_c$ | 0.030 |
| $Nu_b$ | 0.114 |
| $Nu_{b1}$ | 0.147 |
| $Nu_{b2}$ | 0.154 |

To further examine local prediction confidence, prediction distributions for representative test samples were visualized using boxplots, as shown in **Figure 15**. The widths of these distributions remain bounded and consistent, and no pathological variance or extreme outliers were observed. This indicates that the model provides well-calibrated uncertainty estimates and remains stable across most of the feature space.

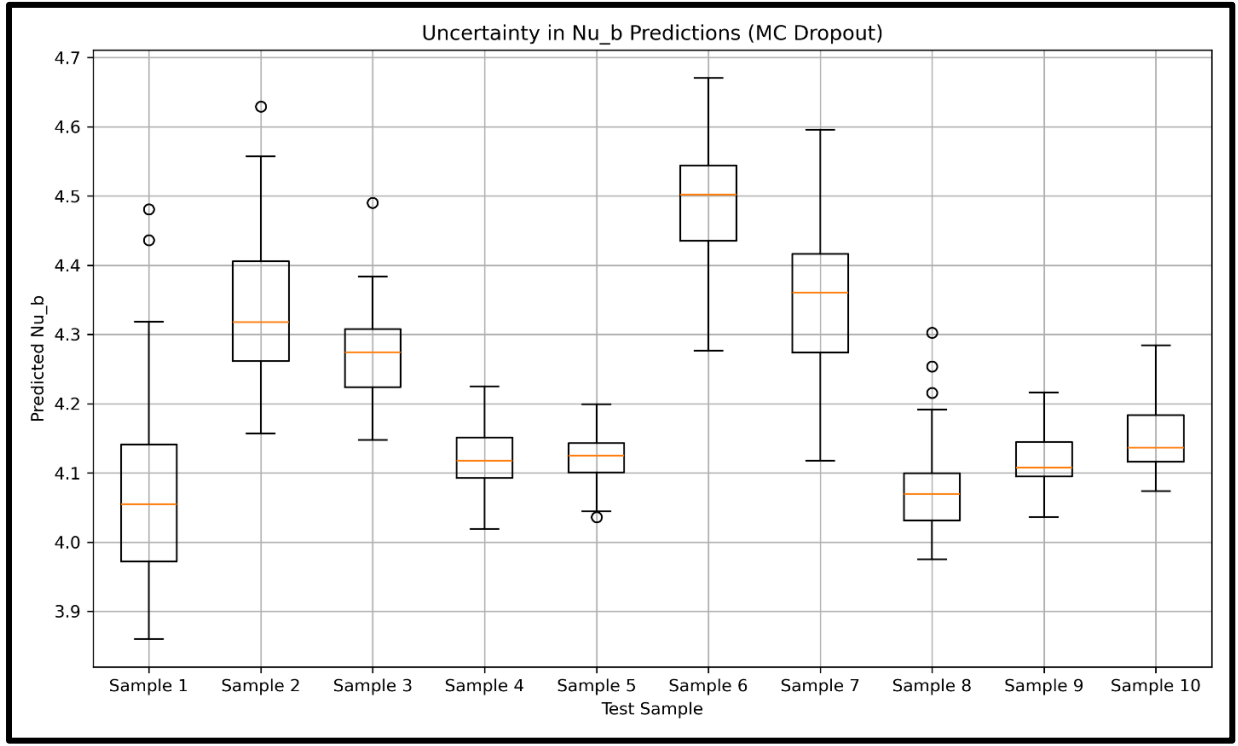

**Figure 15** Uncertainty in $Nu_b$ predictions

Overall, the uncertainty quantification results are consistent with findings from SHAP and permutation importance analyses. Regions associated with stable, globally averaged flow behavior exhibit lower uncertainty, while regions characterized by strong geometric and magnetic interactions show higher uncertainty. This alignment between prediction uncertainty and physical complexity confirms that the neural network not only generalizes well but also provides meaningful confidence estimates, making it suitable for design-oriented and risk-aware thermal modeling applications.

### 4.2.4 Ablation Study

Ablation studies, originally rooted in neuroscience, have become a widely adopted methodology in machine learning to assess the functional contributions of individual components within complex models [31]. In the context of explainable AI, ablation has been extended to input-level perturbations to evaluate the robustness of feature importance estimates and model sensitivity [32]. Ablation studies are commonly used in machine learning to understand how much each input feature contributes to a model's predictive performance. In this study, we performed a systematic feature ablation analysis to evaluate the importance of the seven physical input parameters used to predict the four Nusselt numbers. Each feature was removed one at a time, the neural network was retrained using the remaining inputs, and the resulting predictive performance was evaluated using the coefficient of determination ($R^2$) and root-mean-square error (RMSE).

We report the average R² and RMSE values across the four outputs after each single-feature removal in **Figure 16**. The goal of this analysis was to identify which physical parameters are essential for accurate heat-transfer prediction and which parameters provide only limited independent information. If removing a feature causes a large drop in predictive performance, that feature plays a critical role in the model. Conversely, if performance remains largely unchanged, the removed feature is likely less influential or partially redundant.

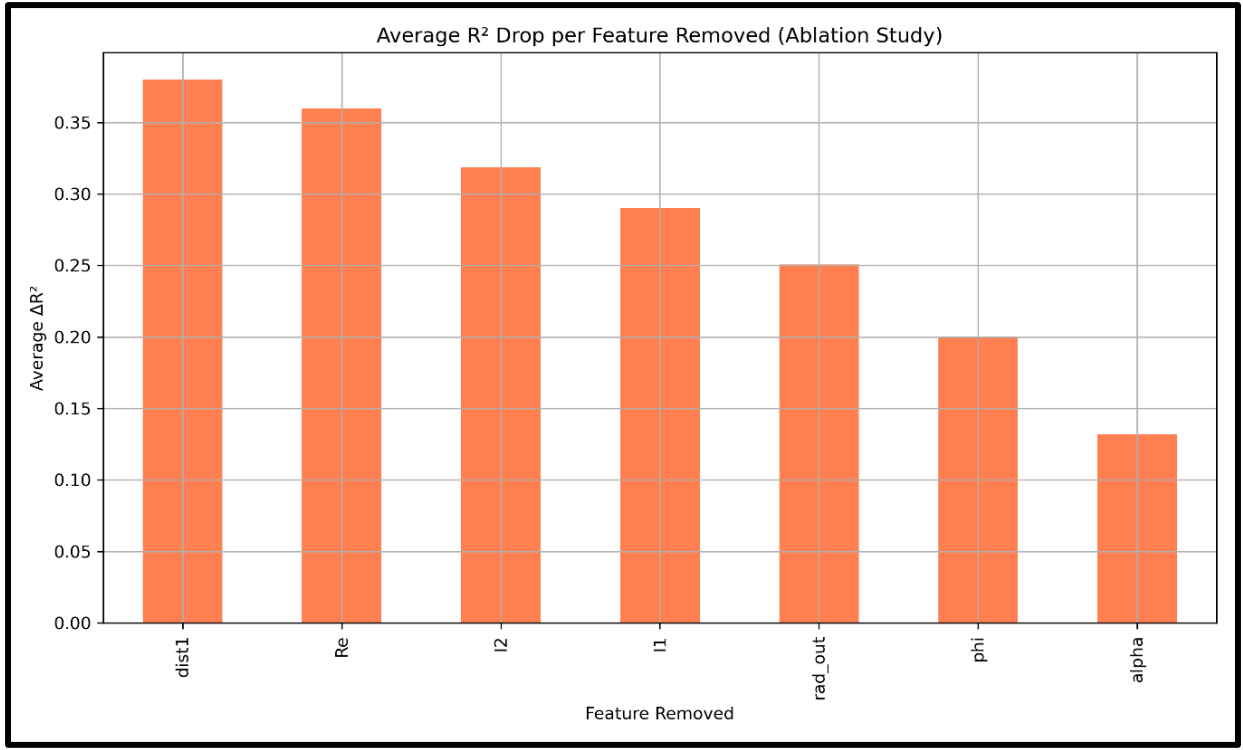

**Figure 16** Ablation Study Bar plot

**Table 5** provides the with summary results for the R² values.

**Table 5** Average R² after removal of each input feature

| Feature Removed | Avg. $R^2$ |
|---|---|
| $dist_1$ | 0.559 |
| $Re$ | 0.579 |
| $I1$ | 0.620 |
| $I2$ | 0.649 |
| $R_o$ | 0.688 |
| $phi$ | 0.739 |
| $alpha$ | 0.807 |

The ablation results show that removing the distance between wires and bend center ($dist_1$) leads to the largest degradation in model performance, with the average $R^2$ dropping below 0.60. This clearly indicates that spatial proximity between the current-carrying wires and the bend is the most important parameter for predicting heat transfer. This finding is consistent with both the SHAP and permutation importance analyses and reflects the strong dependence of magnetic field gradients on distance. The Reynolds number ($Re$) is the second most critical parameter. Its removal also causes a significant decrease in predictive accuracy, confirming that flow inertia strongly influences convective heat transfer in both straight and curved sections of the channel. Together, $dist_1$ and $Re$ form the primary control parameters governing the interaction between magnetic forces and flow dynamics.

Removing the magnetic current inputs ($I_1$ and $I_2$) and the outer bend radius ($R_o$) results in intermediate reductions in performance. This indicates that these parameters play an important but secondary role, primarily by modifying local magnetic forcing and curvature-induced secondary flows. The influence of these features varies across the different Nusselt number outputs, which is consistent with the region-specific trends observed in SHAP and permutation importance results. The ferrofluid volume fraction ($\phi$) produces a smaller but noticeable decrease in model performance when removed. This suggests that while magnetic susceptibility affects heat transfer, its influence is weaker than that of geometry and flow inertia under the conditions studied. Finally, the wire orientation angle ($\alpha$) results in the smallest performance drop, indicating that it contributes the least independent information. This observation aligns with its consistently low ranking in both SHAP and permutation importance analyses.

Overall, the ablation study strengthens confidence in the model by demonstrating consistent feature importance rankings across three independent interpretability methods: SHAP, permutation importance, and feature ablation. The agreement between these methods confirms that the neural network has learned physically meaningful relationships rather than spurious correlations. In particular, the dominant roles of $dist_1$ and $Re$ across all analyses highlight the importance of magnetic field gradients and flow inertia in governing ferrofluid heat transfer in bent channels.

### 4.3 Tertiary performance Evaluation

To complement our quantitative performance metrics and interpretability analyses, we examined the distribution of residual errors (i.e., the differences between predicted and actual values) for all four Nusselt number outputs. Residual analysis serves as a crucial diagnostic tool in regression modeling, revealing potential model miscalibrations, heteroscedasticity, or biases that might not be apparent from metrics like $RMSE$ or $R^2$ alone [22,33].

For each output $Nu_c$, $Nu_b$, $Nu_{b1}$, and $Nu_{b2}$ we generated histograms of the residuals using the test set predictions, as shown in **Figure 17**.

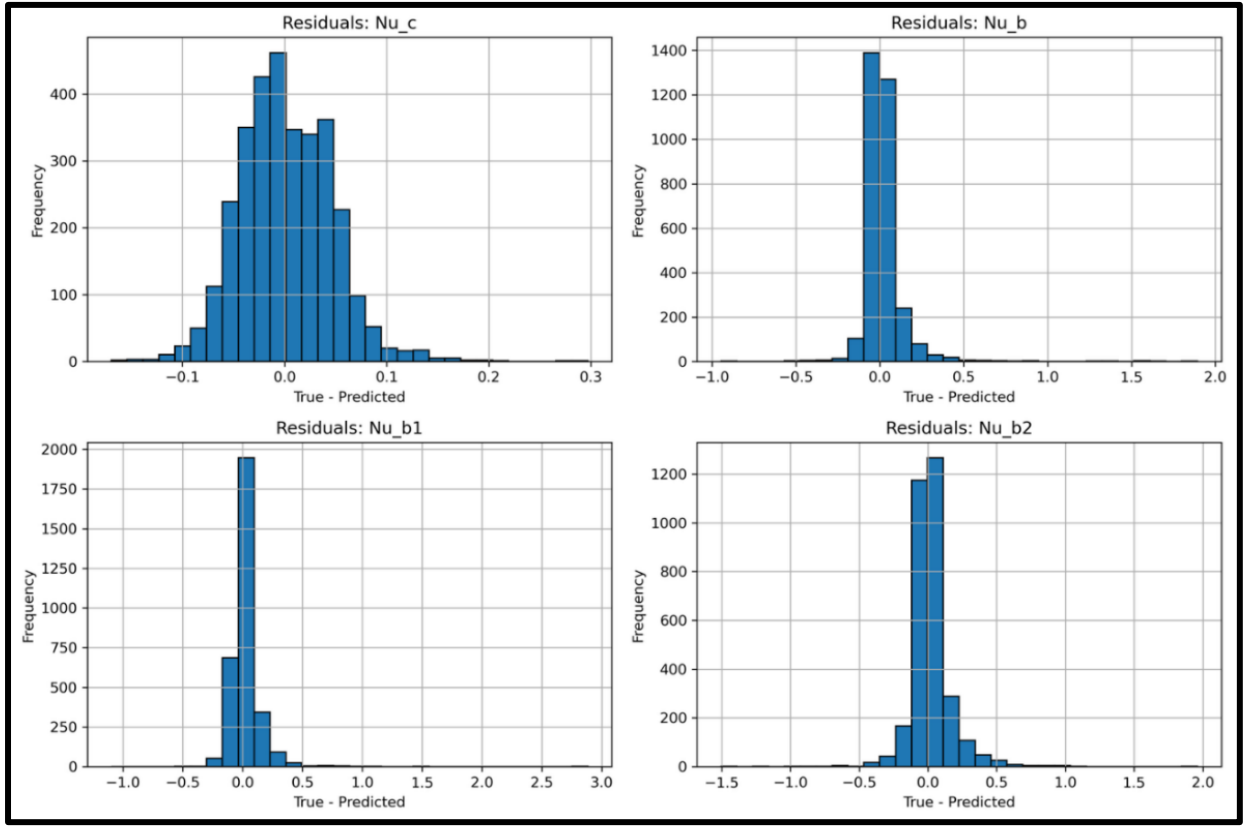

**Figure 17** Residual Error Histograms for Nusselt Numbers

The residuals for the whole-channel Nusselt number, $Nu_c$, formed a narrow, symmetric distribution centered close to zero. This indicates low variance and good model calibration. The shape of the distribution suggests that heat transfer in the straight channel regions is relatively stable and easier for the model to predict consistently.

For the bend-averaged Nusselt number, $Nu_b$, the residual distribution remained largely symmetric but exhibited a broader spread compared to $Nu_c$. This reflects the increased complexity of heat transfer within the curved region, where inertia, curvature-induced secondary flows, and magnetic forcing interact.

The first and second bend-section Nusselt numbers, $Nu_{b1}$ and $Nu_{b2}$, showed slightly heavier tails in their residual distributions. In particular, a mild tendency toward under-prediction was observed. This behavior is consistent with the higher flow variability and sharper thermal gradients present in the bend, especially near geometric transitions. These regions are more sensitive to local changes in magnetic field strength and flow structure, making accurate prediction more challenging.

Importantly, none of the residual distributions showed strong skewness, systematic bias, or increasing variance across the output range. The near-zero mean and overall symmetry of the residuals confirm that the neural network is well calibrated across the range of operating conditions studied. Slightly higher variability appears only in regions where geometric curvature and magnetic effects are strongest, which is physically expected.

Overall, the residual analysis confirms that the model provides stable and reliable predictions across different flow regimes. It also supports the findings from SHAP, permutation importance, and uncertainty quantification by showing that increased prediction uncertainty coincides with physically complex regions of the flow.

## 5. Conclusion

This study proposes a robust and interpretable machine learning framework for predicting convective heat transfer behavior in magnetically influenced ferrofluid flows. Leveraging a fully connected neural network architecture, we successfully modeled four distinct Nusselt number outputs $\boldsymbol{Nu_c}$, $\boldsymbol{Nu_b}$, $\boldsymbol{Nu_{b1}}$, and $\boldsymbol{Nu_{b2}}$ using seven physically meaningful input features that capture geometric, flow, and magnetic parameters.

The model achieved strong predictive performance, with $R^2$ values above 0.97 for three outputs and 0.83 for $Nu_c$, along with good agreement between predicted and actual values. Model transparency was ensured through permutation importance and SHAP analyses, which revealed consistent and physically intuitive feature dependencies. Prediction uncertainty was quantified using Monte Carlo Dropout, showing higher uncertainty in regions of increased flow complexity, particularly within the bend. Ablation studies and residual error analysis further confirmed the robustness of the learned relationships and the reliability of the model.

Overall, the neural network achieved accuracy comparable to traditional ensemble methods such as XGBoost and Random Forests, while offering better scalability and seamless integration with uncertainty estimation. This makes the proposed approach a practical and efficient surrogate for CFD-based heat-transfer modeling.

Future extensions of this work may include spatially resolved predictions using graph neural networks (GNNs), time-dependent modeling under transient magnetic or flow conditions, and application to other multiphase or nanofluid systems. These may include hybrid nanoparticle suspensions, microchannel heat sinks, and adaptive cooling technologies in electronics and biomedical applications.

This work demonstrates that physics-informed machine-learning models, when combined with rigorous interpretability and validation techniques, can provide accurate, reliable, and scalable solutions for complex thermofluid problems.

# Appendix A

Table 6 Feature-wise Mean Importance & Std Dev of all Output Variables

| Target | Feature | Mean Importance | Std Dev |
|---|---|---|---|
| $Nu_c$ | $dist_1$ | 0.012633 | 0.000333 |
| $Nu_c$ | $I_2$ | 0.010076 | 0.000464 |
| $Nu_c$ | $I_1$ | 0.009052 | 0.000462 |
| $Nu_c$ | $Re$ | 0.005164 | 0.000145 |
| $Nu_c$ | $\phi$ | 0.004703 | 0.000251 |
| $Nu_c$ | $\alpha$ | 0.002927 | 0.000085 |
| $Nu_c$ | $R_o$ | 0.001732 | 0.000155 |
| $Nu_b$ | $Re$ | 0.562302 | 0.030282 |
| $Nu_b$ | $dist_1$ | 0.457766 | 0.018422 |
| $Nu_b$ | $R_o$ | 0.383974 | 0.033715 |
| $Nu_b$ | $I_2$ | 0.382223 | 0.023461 |
| $Nu_b$ | $I_1$ | 0.321742 | 0.020822 |
| $Nu_b$ | $\phi$ | 0.252158 | 0.018592 |
| $Nu_b$ | $\alpha$ | 0.116476 | 0.006746 |
| $Nu_{b1}$ | $Re$ | 1.006595 | 0.042321 |
| $Nu_{b1}$ | $dist_1$ | 0.584305 | 0.031588 |
| $Nu_{b1}$ | $I_1$ | 0.508455 | 0.039738 |
| $Nu_{b1}$ | $I_2$ | 0.413207 | 0.026712 |
| $Nu_{b1}$ | $\phi$ | 0.369336 | 0.028848 |
| $Nu_{b1}$ | $R_o$ | 0.233196 | 0.019385 |
| $Nu_{b1}$ | $\alpha$ | 0.199185 | 0.012247 |
| $Nu_{b2}$ | $R_o$ | 0.697014 | 0.039024 |
| $Nu_{b2}$ | $I_2$ | 0.586159 | 0.027246 |
| $Nu_{b2}$ | $Re$ | 0.534364 | 0.026009 |
| $Nu_{b2}$ | $dist_1$ | 0.517245 | 0.024107 |
| $Nu_{b2}$ | $I_1$ | 0.364470 | 0.026334 |
| $Nu_{b2}$ | $\phi$ | 0.269596 | 0.019138 |
| $Nu_{b2}$ | $\alpha$ | 0.146711 | 0.006413 |